\documentclass[twocolumn,prd,nofootinbib,showpacs,floatfix,preprintnumbers]{revtex4}
\usepackage{epsfig}
\usepackage[dvips]{color}

\def\alt{\raise0.3ex\hbox{$\;<$\kern-0.75em\raise-1.1ex\hbox{$\sim\;$}}}
\def\agt{\raise0.3ex\hbox{$\;>$\kern-0.75em\raise-1.1ex\hbox{$\sim\;$}}}

\definecolor{Black}{named}{Black}
\definecolor{Red}{named}{Red}

\newcommand{\bw}{\begin{widetext}}
\newcommand{\ew}{\end{widetext}}

\begin{document}

\title{Ultra-high energy Neutrinos from Centaurus A and the Auger hot spot}
\author{A.~Cuoco$^{1}$,  S.~Hannestad$^{1}$}
\affiliation{$^1$Department of Physics and Astronomy, University
of Aarhus, Ny Munkegade, Bygn. 1520 8000 Aarhus Denmark }

\date{\today}

\begin{abstract}
The Pierre Auger collaboration has reported a correlation between
Ultra-High Energy Cosmic Rays (UHECR) and nearby Active Galactic
Nuclei (AGNs) within $\sim$75 Mpc. Two of these events fall within 3
degrees from Centaurus A, the nearest AGN, clearly suggesting that
this object is a strong UHECR emitter. Here we pursue this
hypothesis and forecast the expected rate of ultra-high energy
neutrinos in detectors like IceCube. In our baseline model we find a
rate of $\sim$ 0.4--0.6 yr$^{-1}$ events above a threshold of 100
TeV, the uncertainty of which is mainly related to the poor
knowledge of the physical parameters of the source and on the
details of the model. This situation will improve with detailed high
energy gamma ray measurements of Cen A by the upcoming GLAST
satellite. This would make Cen A the first example where the
potential of high energy multi-messenger astronomy is finally
realized.
\end{abstract}


\maketitle

The field of UHECR physics has probably taken a major step forward
with the recent detection by the Pierre Auger Observatory of a
spatial correlation between the highest energy cosmic ray events and
nearby AGNs \cite{Abraham:2007bb}. 20 out of 27 events with energies
above $\simeq60$ EeV correlate with a nearby AGN within a radius of
3.1$^\circ$. Furthermore, 5 out of the 7 non-correlating events lie
along the galactic plane where the AGN catalogues are incomplete and
the largest magnetic deflections are expected.

With such support for the hypothesis that AGNs are the main
emitters of UHECRs it is timely to explore the consequences for
other areas of high energy astrophysics. We will focus on the
connection between UHECRs and neutrinos \cite{reviews} and
investigate the possibility for detecting the neutrinos associated
with the UHECR acceleration in AGNs. From the experimental point
of view the detection of UHE diffuse and point source neutrinos in
the km$^3$ IceCube detector is promising \cite{:2007td} and the
AMANDA collaboration has already reported interesting limits
\cite{Ackermann:2007km}.

Roughly 10 events are concentrated in the Centaurus direction, a
region with a high density of AGNs, constituting a hot spot in the
Auger UHECR map. Our focus will be on Centaurus A (Cen A) as a
case study (see also \cite{Anchordoqui:2004eu}). Two events fall
near this galaxy, suggesting that it could be the first identified
UHECR source. Indeed, Cen A, the nearest AGN at a distance of only
$\sim 4$ Mpc \cite{Israel:1998ws}, has long been considered as a
prime UHECR source candidate \cite{Torres:2004hk}. The problem of
predicting the neutrino flux from Cen A is similar to the attempts
to relate the observed UHECRs diffuse flux with a prediction, or
at least with an upper bound, for the diffuse UHE neutrino flux
\cite{Waxman:1998yy,Bahcall:1999yr,Mannheim:1998wp}. We will
employ basically the same approach for the case of Cen A with the
help of the available data on its spectral distribution. Various
models of neutrino emission from AGNs have been discussed in the
past (see for example \cite{Stecker:1991vm}). A recent update has
been considered in \cite{Kachelriess:2006fi,Kachelriess:2007tr}.
We describe our model in more detail below.

\section{The Auger Flux}
The expected number of events in the Auger
array can be calculated starting from the total integrated exposure.
The auger group reports $\Xi=9000$ km$^2$ yr~sr at present
\cite{Abraham:2007bb}. For point sources we need the exposure per
steradian given by $\Xi/\Omega_{60}$ where $\Omega_{60}=\pi$ sr is
the Auger field of view corresponding to 60 degrees as maximum
zenith angle. In addition the relative exposure $\omega(\delta)$ is
required, weighting a source with declination $\delta$ for the
effective observation time. $\omega(\delta)$ is parameterized
according to \cite{Sommers:2000us} using $\theta=-35^\circ$ for the
Auger declination and normalizing to 1 the maximum. Assuming a power
law shape for the energy spectrum $F=F_0 (E/E_0)^{-\alpha}$ we get
\begin{equation}
    N=  F_0 \ \frac{\Xi \ \omega(\delta_s)\ E_0}{\Omega_{60}\ (\alpha-1)} \left( \frac{E_c}{E_0} \right)^{1-\alpha}
\end{equation}
where $E_c$ is the threshold energy, or, equivalently
\begin{equation}\label{UHEflux}
    F = \frac{N \ \Omega_{60} (\alpha-1)}{\Xi \ \omega(\delta_s) \ E_0}
    \left( \frac{E_c}{E_0}  \right)^{\alpha-1}
      \left( \frac{E}{E_0} \right)^{-\alpha}.
\end{equation}
For the case of Cen A we have $N=2$ events above  a threshold
$E_c=60$ EeV with a source declination $\delta_s\simeq -47^\circ$
and relative exposure $\omega(\delta_s)\simeq 0.64$ which gives
\begin{equation}
    F  \simeq  1.95 \
  \left( \frac{E}{\rm EeV} \right)^{-2.7} \frac{1}{\rm{km^2 \ yr \ EeV}}
\end{equation}
or $E^3F\simeq 6\times10^{22}(E/ \rm EeV)^{0.3}\ eV^2/\rm{m^2 \ s}$.
The uncertainty on the flux estimate is roughly $\sqrt2/2\sim70\%$
from Poisson statistics. The intrinsic slope with just two events is
very uncertain and we thus use $\alpha=2.7$ as seen in the diffuse
UHECR flux just before the GZK cutoff assuming it as generally
representative of the typical UHECR emitter. The uncertainty in the
Cen A flux is therefore significant, but the situation is expected
to improve as more statistics is collected by the Auger array
allowing, in principle, to constrain the spectral index directly
from the data. Further, once the source is clearly identified, also
lower energy events can be used to reconstruct the spectrum, despite
the larger magnetic deflections. We will see in the following,
however, that the main source of uncertainty in the $\nu$ flux is
the AGN modeling rather than the UHECR flux uncertainty. An
additional uncertainty is related to the possible systematic error
on the absolute energy scale of Auger of up to 30\%. An independent
calibration is in principle possible exploiting the dip feature
present in the UHECR spectrum at $\sim 10^{18}$ eV \cite{dip}. This
gives $E_c=80$ EeV and a flux roughly a factor $\sim 1.6$ higher.
Below we refer to this as the ``dip'' energy scale.

Dropping the factor $\Omega_{60}/\omega(\delta_s)$ we can use
Eq.~(\ref{UHEflux}) also to estimate the diffuse UHE flux. Using
$N$, $E_c$, $\Xi$ from \cite{Roth:2007in} we find $F\simeq 21 \left(
E/\rm EeV \right)^{-2.7} 1/\rm{km^2 \ yr \ sr \ EeV}$ or $E^3F\simeq
2\times10^{24}(E/ \rm EeV)^{0.3}\ eV^2/\rm{m^2 \ s \ sr}$, in good
agreement with the Auger estimate itself \cite{Roth:2007in}.

\section{AGN modeling and $\nu$ Flux}\label{model}

To relate the expected neutrino flux to the observed CR flux several
assumptions and an underlying model are unavoidably required. The
usual scenario assumes that protons are shock-accelerated to
ultra-high energies and then interact with ambient radiation or
matter producing secondary neutrons and pions with an associated
flux of gammas and neutrinos. If we assume that protons are
magnetically confined in the source so that only neutrons can escape
producing the observed flux of UHECRs, a direct link between
neutrinos and UHECRs, or more generally an upper limit, is possible
\cite{Waxman:1998yy,Bahcall:1999yr,Mannheim:1998wp}. In the
following we will use this hypothesis. The presence of particles
accelerated up to $10^{20}$ eV indeed implies a magnetic field
generally strong enough to confine the particles themselves.
Furthermore, even if the protons can finally diffuse out of the
source, the acceleration region is generally expanding so that
adiabatic losses limit the maximum attainable energies
\cite{Rachen:1998fd}.

UHE protons can interact in the source both with matter through
$p+p \to p(n)+\pi$s or with the local radiation field through
$p+\gamma \to p(n)+\pi^{0(+)}$. Subsequent pion decay then
produces photons and neutrinos. The collisionless conditions
required for efficient shock-acceleration imply a relatively low
matter density so that at ultra-high energies the dominant
neutrino production channel is generally the photo-hadronic one.
We will thus focus on this process, using the Monte Carlo code
\verb"SOPHIA" \cite{Mucke:1999yb} to simulate the interactions of
protons in the Cen A radiation field and to normalize the relative
yields and multiplicities of the secondary particles per
interaction. These processes occur mainly close to threshold and
produce only 1-2 pions per interaction. $p$--$p$ interactions,
instead, although disfavored, have quite higher ($\agt 10$) pions
multiplicities  \cite{Razzaque:2002kb} and thus would also give
higher neutrino multiplicities. Finally, we will neglect muons and
pions synchrotron losses in the source MF that however are
expected to affect the $\nu$ flux only at the highest energies
(see \cite{Kachelriess:2007tr,Lipari:2007su} for a throughout
discussion on the role of MFs).

\begin{figure}[tbp]
\begin{center}
\epsfig{file=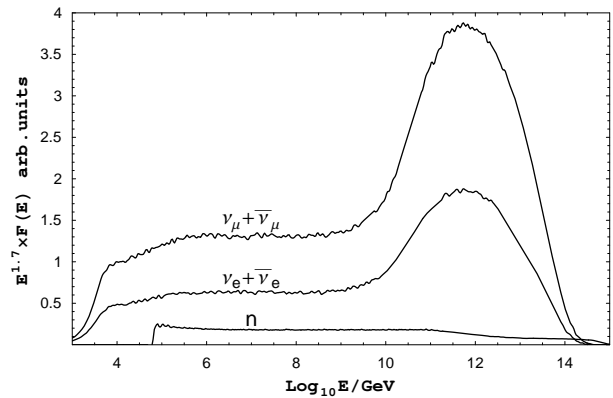,width=0.99\columnwidth,angle=0}
\end{center}
\vspace*{-0.7cm} \caption{Resulting neutron,
$\nu_{\mu}$+$\bar{\nu}_{\mu}$ and $\nu_{e}$+$\bar{\nu}_{e}$
spectra for an $E^{-1.7}$ and $10^5$$<$$E/\rm{GeV}$$<$$10^{15}$
proton injection  spectrum interacting on the photon field of Cen
A simulated  with \texttt{SOPHIA} in the optically thin source
limit.}\label{fig1} \vspace*{-0.0cm}
\end{figure}

The final neutrino yield has also a dependence on the Spectral
Energy Distribution (SED) of the Cen A radiation field. In
particular, if numerous enough ambient high-energy (x-ray/gamma)
photons and thus high CM energies are available, in principle the
multi-pion production channels can be activated and an higher
neutrino multiplicity per interaction can be achieved with respect
to the expected low near-threshold yield. We consider especially
$\nu$ production in the nucleus for which detailed SED information
is available. The jet case is more uncertain and will be discussed
briefly.  A compilation of measurements of the Cen A nucleus is
reported in \cite{Chiaberge:2001ek}. The SED has an approximate
double peak structure with an infrared peak and a second soft-gamma
peak. Upper limits in the TeV region at the level of \mbox{few \%}
of the Crab flux has been reported by the current generation of VHE
cherenkov telescopes \cite{Aharonian:2005ar,Kabuki:2007am} .  We
adopt for the photon number density the crude approximation of a
broken power law $n(\epsilon)\propto \epsilon^{-1.9}$ over the range
$0.001\ \rm{eV}<\epsilon<100\ \rm{MeV}$ and $n(\epsilon)\propto
const.$ for $\epsilon<0.001\ \rm{eV}$, accurate enough to determine
the actual neutrino production regime. We show in Fig.~\ref{fig1}
the resulting neutron and neutrino spectra for an $E^{-1.7}$ proton
injection spectrum interacting on the photon field of Cen A
simulated  with \texttt{SOPHIA} in the optically thin source limit,
in which multiple proton-photon interactions have a negligible role.
Indeed, it can be seen from the figure that the interactions occur
near threshold for the whole relevant energy range, while only at
very high, uninfluential, energies \mbox{$E_\nu>10^{10}$ GeV} we
enter in the higher neutrino production regime. We therefore use the
relevant average quantities valid in the $E_\nu<10^{10}$ GeV range,
namely: $\left<\xi_\nu\right>\simeq0.1$, $\left<\xi_n\right> \simeq
0.5$, as the fraction of proton energy transferred respectively to
the neutrino ($\nu_e+\nu_\mu+$ their antiparticles) and neutron
component per interaction, and $\eta_{\nu
n}=\left<E_\nu\right>/\left<E_n\right>\simeq 0.04$ for the average
neutrino to neutron energy.  The final neutron and neutrino flux are
thus simply related via
\begin{equation}\label{NuFluxNFlux}
    F_\nu(E) = \frac{\left<\xi_\nu\right>}{\left<\xi_n\right> \eta_{\nu n}^2}
                     F_n(E/\eta_{\nu n})
\end{equation}
where the factor $\left<\xi_\nu\right>/\left<\xi_n\right> \eta_{\nu
n} \simeq 5$ gives the average mean neutrino/neutron multiplicity.
Notice that the multiplicity is $>3$ implying that, even near
threshold, more than one pion is on average produced per
interaction. Our estimates are  in fair agreement with
ref.~\cite{Mucke:1999yb} to which we address the reader for a more
detailed discussion of the various neutrino production regimes in
photo-hadronic processes.

If, on the other hand, the production site is located in the Cen A
jet rather than in the nucleus, a softer, more \mbox{x-ray}
populated, $n(\epsilon)$ photon spectrum can be possible
\cite{Hardcastle:2003ye} and an higher neutrino multiplicity is
achievable. However, in this case the site of acceleration would
be the shock regions/hot spots in the jet, where a large
scattering in the spectral indexes is observed
\cite{Hardcastle:2003ye} so that a firm prediction is hard to
establish. We also remark that Cen A is classified as a misaligned
BL Lac with its jet pointing
$20^\circ$-$40^\circ$\cite{Horiuchi:2005kz} away from our line of
sight so that the relativistic boosting effect should not play a
major role if the UHECRs come from the jet.

\begin{figure}[tbp]
\vspace{-0.9pc}
\begin{center}
\epsfig{file=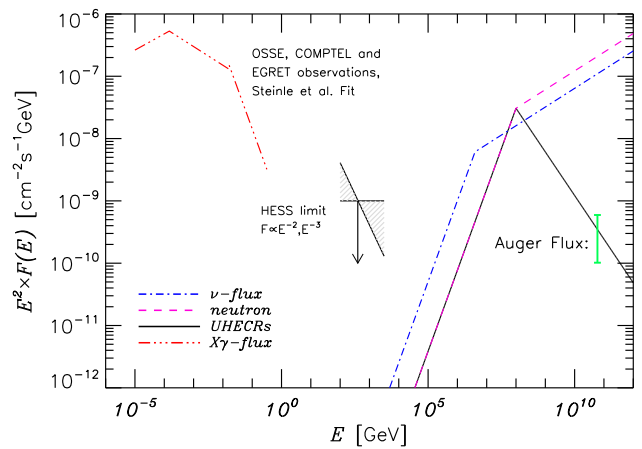,width=1.05\columnwidth,angle=0}
\end{center}
\vspace*{-0.7cm} \caption{Final neutron, UHECRs and total neutrino
spectrum in the model of this work, normalized to the Auger
observation at $E\simeq 60$ EeV.  Also shown is the fit to the MeV
soft-gamma observations \cite{Steinle:1998yr,Kinzer:1995yr} and the
HESS upper limit around $\sim 1$ TeV \cite{Aharonian:2005ar}. The
Cen A UHE-$\gamma$ curve, not shown for clarity, lies close to the
neutrino curve. }\label{fig2} \vspace*{-0.0cm}
\end{figure}

To extrapolate the neutrino flux from the CR flux to PeV energies
further modeling of the internal acceleration mechanism of the
source is required. Extrapolating the $E^{-2.7}$ spectrum to very
low energies is clearly unrealistic and in fact several breaks in
the slope of the energy spectrum with subsequent steepening as the
energy increases are predicted. We follow
\cite{Mannheim:1998wp,Rachen:1998fd} for the modeling of these
breaks in the neutron spectrum and to relate it to the observed
UHECRs spectrum. We thus consider a scenario in which an ambient
proton spectrum $\propto E_p^{-1.7}$ interacts with the low energy
radiation field producing neutrons that, escaping from the source,
decay into the observed CRs spectrum. The proton injection index 1.7
is in general agreement with the typical value $\approx 2$ expected
from shock acceleration and it is chosen in such a way that the CR
spectral index matches the value 2.7 at Ultra High energies (see
below). Although the details are generally quite model dependent two
clear breaks are predicted in neutron/CR spectrum in the highest
energy regime. In the first the spectrum steepens by one power when
the pion production process becomes efficient while a second one
power steepening is predicted when the source becomes optically
thick to photo-hadronic interactions. The two breaks are generally
close so we assume a single break at the energy $E_b$. The resulting
UHECR spectrum is then
\begin{equation}
    F_{\rm CR}(E) \propto \left\{
\begin{array}{cc}\nonumber
  E_b^{-2} E^{-0.7} &   (E<E_b),       \\ \nonumber
  E^{-2.7}          &   (E>E_b).
\end{array}
\right.
\end{equation}
There are thus three species in the model with different energy
spectra: underlying, not directly observable protons, with an
injection spectrum $\propto E_p^{-1.7}$, photo-produced underlying
neutrons, with a spectrum $\propto E_n^{-0.7}$ before the break and
$\propto E_n^{-1.7}$ after the break thus following the proton
spectrum, and escaping ``physical'' neutrons, with spectrum further
showing one power steepening above the break, $\propto E_{\rm
CR}^{-2.7}$. The escaping neutrons then decay back into protons far
from the source constituting the final CR spectrum whose UHE tail is
observed in Auger. The various nuclear species spectra and the
neutrino flux are shown in Fig.~\ref{fig2}. The proton injection
flux is not shown for clarity. Notice that the neutrino spectrum is
in general supposed to follow the \emph{underlying}, neutron
spectrum via Eq.(\ref{NuFluxNFlux}), i.e. unattenuated by pion
losses and with a behavior $\propto E^{-1.7}$ above the break
energy. We will see however that the final expected rate of
neutrinos is insensitive to the exact behavior of the spectrum above
the break. It also does not depend crucially on the slope below the
break as long as it remains in the range between 1-2. The main
parameter determining the neutrino rate is the actual break energy
itself. To estimate this we use observations of the gamma spectrum
from Cen A.

The most important point to take into account to this aim is that
gamma photons interact via pair production with the same low energy
photon background relevant to the photo-hadronic interactions  so
that a break in the gamma spectrum can be related to a break in the
UHECR spectrum. In particular, considering the ratio of the related
cross sections and inelasticities it can be seen that
\cite{Mannheim:1998wp,Bahcall:1999yr}
$\tau_{n\gamma}(E_n)/\tau_{\gamma\gamma}(E_\gamma) \simeq
4\times10^{-9} E_n/E_\gamma$, where $\tau_i$ is the optical depth of
the related process. When the source becomes optically thick to both
processes (i.e. $ \tau_{n\gamma}=\tau_{\gamma\gamma}=1$) we thus
have the relation \mbox{$E_{b\gamma} \simeq 4 \times10^{-9}\
E_{bn}$} between the neutron and gamma break energies. Thus,
differently from the determination of the neutrino multiplicity, the
exact shape of the spectrum is in this case crucial. Cen A
observations in the gamma band
\cite{Steinle:1998yr,Kinzer:1995yr,Sreekumar:1999xw} show several
breaks in the range \mbox{100 keV-100 MeV} with a photon spectral
index $\alpha\simeq 1.7$ for $E\alt 200$ keV and $\alpha\simeq 3.0$
for $10 \alt E \alt100$ MeV \cite{Steinle:1998yr}. The exact energy
of the highest break, however, is time dependent due to the
intrinsic variability of the source. Further, the observations
become photon limited above 200 MeV \cite{Sreekumar:1999xw}, making
unclear if a further relevant spectral steepening is present above
this energy. To be conservative we assume a gamma break at the
highest detected energy bin found by EGRET at $E_{b\gamma}\simeq
200$ MeV, implying $E_{bn}\sim 10^8$ GeV. Anyway, given the
importance of this parameter, that basically determines the
normalization of the neutrino flux, in the following we also analyze
the effect of a different choice for $E_{bn}$. We show in
Fig.\ref{fig2} our final neutrino, neutron and UHECR spectra
together with a fit to the MeV soft-gamma observations
\cite{Steinle:1998yr,Kinzer:1995yr} and the HESS upper limit around
$\sim 1$ TeV \cite{Aharonian:2005ar}.

We finally comment on the hadronic associated gamma flux expected to
accompany the  CR and neutrino fluxes (see also
ref.~\cite{Kachelriess:2008qx} for a specific analysis of the
issue). The $\left<\xi_\gamma\right>$ and $\eta_{\gamma n}$ factors
are indeed very similar to the neutrino case so that the neutrino
and gamma fluxes are predicted to be very close in shape and
normalization. However, while neutrinos leave the sources just after
production, gammas are subject to further processing through the
development of an electro-magnetic cascade that depletes the high
energy photon tail producing sub-TeV photons. Pair production in the
low energy photon field of the source and electron synchrotron
losses need thus to be taken into account for a prediction of the
observable gamma flux. This, in turn, require further modeling of
the source adding further uncertainties in the predictions.

\section{Event rate in a Neutrino Telescope}
Pion decay yields the flavor ratio
$\nu_e\,$:$\,\nu_\mu\,$:$\,\nu_\tau=1\,$:$\,2\,$:$\,0$. However, due
to oscillations, we expect the ratio
$\nu_e\,$:$\,\nu_\mu\,$:$\,\nu_\tau=1\,$:$\,1\,$:$\,1$ at Earth. At
energies $E_\nu\agt 100$ TeV, neutrino telescopes are fully
efficient both to tracks from charged current generated $\mu$'s and
to showers from $\nu_e$ events. $\tau$ leptons from $\nu_\tau$
interactions are expected to be detected both as showers near
$E=100$ TeV and as tracks in the higher energy range. For southern
hemisphere detectors, like IceCube, neutrino events from Cen A are
down-going. In particular, Cen A, with a declination $\delta\simeq
-47^\circ$ appears at a zenith angle of $43^\circ$ in the IceCube
field of view.

Regarding the background, for $E_\nu\agt 100$ TeV the atmospheric
neutrino flux is negligible while the residual background of
atmospheric muons has a quite steep spectrum rapidly decreasing with
energy. For simplicity we assume full efficiency for UHE neutrino
detection, while a more careful evaluation would require a detailed
Monte Carlo simulation. The final background is thus constituted by
the UHE diffuse $\nu$-flux itself whose relevance is linked to the
detector angular resolution which at these energies is $\sim$ few
degrees \cite{:2007td}. We will limit ourselves to estimate the
expected $\nu$-flux while assessing the corresponding statistical
significance will eventually rely on the measured diffuse flux
normalization. To calculate the event rate in a km$^3$ detector like
IceCube we assume for showers an effective volume of $V_{\rm eff}=2$
km$^3$. For track events the effective volume could be much higher
due to the muon and tau range. However, for a zenith angle of
$43^\circ$ the available overburden is limited. We therefore
conservatively use the same $V_{\rm eff}$ also for tracks. From the
effective volume the expected event rate is
\begin{equation}\label{eventsrate}
    \mathcal{N}=N_A \rho V_{\rm eff} \int_{E_{\rm th}}^{+\infty} \!\!\!\!
dE_\nu   \ \sigma^{CC}_{\nu N} \ F_\nu(E_\nu)
\end{equation}
where $N_A$ is the avogadro number, $\rho$ the density of the
target material (ice in this case), $E_{\rm th}=100$ TeV is the
threshold energy and $\sigma^{CC}_{\nu
N}=6.78\times10^{-35}(E_\nu/\rm{TeV})^{0.363}$ cm$^{2}$ is the CC
cross section \cite{Gandhi:1998ri}.

Using the neutrino flux from the previous section, gives
$\mathcal{N}\simeq 0.35$ yr$^{-1}$ or $\mathcal{N}\simeq 0.56$
yr$^{-1}$ for the dip calibrated energy. Thus the conclusion is that
IceCube should collect $\mathcal{O}$(few) events from Cen A in 5
years. This could be enough for a confident detection of the source
if the diffuse UHE $\nu$-background is not too high. However,
although the detection of Cen A may be challenging we can in
principle extend the analysis to the whole Auger hot spot assuming
that the related UHECRs emitters share the characteristics of Cen A.
This results in a rate $\mathcal{N}\simeq 2$ yr$^{-1}$ in a region
of radius about 10 degrees centered on Cen A which should be evident
after a few years of observations. As anticipated the estimated
event rate does not dependent crucially on the neutrino flux slope
while it is very sensitive to the break energy, $E_b$, which
determines the relation between the neutrino normalization and the
UHECR normalization as inferred from Auger. In Table \ref{tab:1} we
show the scatter in the values of $\mathcal{N}$ for different values
of $E_b$ and for the Auger and ``dip'' energy scales. An order of
magnitude variation in $\mathcal{N}$ is in principle possible if the
value of $E_b$ differs correspondingly by one order of magnitude.
Clearly further observations in the gamma band would be desirable to
have a more robust estimate of $E_b$. Fortunately the situation is
expected to improve with the launch of the GLAST satellite that
should provide high quality data up to GeV energies and  possibly
beyond. Also, deeper observations from Cherenkov telescopes in the
TeV range would contribute to improve the picture.

\begin{table}[!t]
\begin{tabular}{|c||c|c|c|}
\hline $E_c$ & $E_b=10^7$ GeV & $E_b=10^8$ GeV & $E_b=10^9$ GeV
\\ \hline
 \hline $60$ EeV & 6.7 yr$^{-1}$ & 0.35 yr$^{-1}$ & 0.016 yr$^{-1}$\\
 \hline $80$ EeV & 11.0 yr$^{-1}$ & 0.56 yr$^{-1}$ & 0.026 yr$^{-1}$\\
\hline
\end{tabular}
\caption{Cen A event rate in IceCube for various break energies
$E_b$ and for the Auger and ``dip'' energy scales.\label{tab:1}}
\vspace{-0pc}
\end{table}

Despite the uncertainties the prospect of neutrino detection from
Cen A and its surroundings are quite promising with the exciting
possibility to perform true multi-messenger astronomy, observing
for the first time a source in UHECRs, neutrinos and
$\gamma$-rays. This would also allow for detailed studies of the
source acceleration mechanism \cite{Anchordoqui:2004eb} and, if
flavor tagging can be achieved, neutrino  exotic properties could
be tested \cite{Lipari:2007su,Serpico:2005sz}.

We conclude by commenting on a puzzling aspect of the Auger data
\cite{Gorbunov:2007ja}: Although many AGNs lie in the direction of
the Virgo cluster, no events are detected. Although the statistics
is low and this could be an exposure effect it is intriguing to
notice that the issue can be settled by observations of the
associated neutrino emission in IceCube.

{\it Acknowledgments ---} We thank F.~Halzen, P.~D.~Serpico and
G.~Miele for valuable comments on the manuscript. Use of the
publicly available \verb"SOPHIA" \cite{Mucke:1999yb} code is
acknowledged. \vspace{-1pc}



\begin{thebibliography}{00}

\bibitem{Abraham:2007bb}
  J.~Abraham {\it et al.},
  Science {\bf 318} (2007) 939.
  J.~Abraham {\it et al.},
  Astropart.\ Phys.\  {\bf 29} (2008) 188.




\bibitem{reviews}
  T.~K.~Gaisser, F.~Halzen and T.~Stanev,
  Phys.\ Rept.\  {\bf 258} (1995) 173
  [Erratum-ibid.\  {\bf 271} (1996) 355].
  F.~Halzen and D.~Hooper,
  Rept.\ Prog.\ Phys.\  {\bf 65} (2002) 1025.
  J.~G.~Learned and K.~Mannheim,
  Ann.\ Rev.\ Nucl.\ Part.\ Sci.\  {\bf 50} (2000) 679.



\bibitem{:2007td}
  The IceCube Collaboration, ``Contributions to the 30th International Cosmic Ray Conference (ICRC 2007),''
  arXiv:0711.0353 [astro-ph].


\bibitem{Ackermann:2007km}
  M.~Ackermann  [The IceCube Collaboration],
  Astrophys.\ J.\  {\bf 675} (2008) 1014.


\bibitem{Anchordoqui:2004eu}
  L.~A.~Anchordoqui et al.,
  Phys.\ Lett.\  B {\bf 600} (2004) 202.
  F.~Halzen and A.~O'Murchadha,
  arXiv:0802.0887.


\bibitem{Israel:1998ws}
  F.~P.~Israel,
   Astron.\ Astrophys.\ Review {\bf 8} (1998) 237-278.





\bibitem{Torres:2004hk}
  D.~F.~Torres and L.~A.~Anchordoqui,
  Rept.\ Prog.\ Phys.\  {\bf 67} (2004) 1663;
  L.~Anchordoqui {\it et al.},
  Int.\ J.\ Mod.\ Phys.\  A {\bf 18} (2003) 2229.






\bibitem{Waxman:1998yy}
  E.~Waxman and J.~N.~Bahcall,
  Phys.\ Rev.\  D {\bf 59} (1999) 023002.

\bibitem{Bahcall:1999yr}
  J.~N.~Bahcall and E.~Waxman,
  Phys.\ Rev.\  D {\bf 64} (2001) 023002.

\bibitem{Mannheim:1998wp}
  K.~Mannheim, R.~J.~Protheroe and J.~P.~Rachen,
  Phys.\ Rev.\  D {\bf 63} (2001) 023003.

\bibitem{Stecker:1991vm}
  F.~W.~Stecker {\it et al.},
  Phys.\ Rev.\ Lett.\  {\bf 66} (1991) 2697
  [Erratum-ibid.\  {\bf 69} (1992) 2738];
  F.~W.~Stecker,
  Phys.\ Rev.\  D {\bf 72} (2005) 107301;
  F.~Halzen and E.~Zas,
  Astrophys.\ J.\  {\bf 488} (1997) 669;
  J.~Alvarez-Muniz and P.~Meszaros,
  Phys.\ Rev.\  D {\bf 70} (2004) 123001;
  A.~P.~Szabo and R.~J.~Protheroe,
  Astropart.\ Phys.\  {\bf 2} (1994) 375;
  K.~Mannheim,
  Astropart.\ Phys.\  {\bf 3} (1995) 295;
  A.~Y.~Neronov and D.~V.~Semikoz,
  Phys.\ Rev.\  D {\bf 66} (2002) 123003.

\bibitem{Kachelriess:2006fi}
  M.~Kachelriess and R.~Tomas,
  Phys.\ Rev.\  D {\bf 74} (2006) 063009.

\bibitem{Kachelriess:2007tr}
  M.~Kachelriess, S.~Ostapchenko and R.~Tomas,
  Phys.\ Rev.\  D {\bf 77} (2008) 023007






\bibitem{Sommers:2000us}
  P.~Sommers,
  Astropart.\ Phys.\  {\bf 14} (2001) 271.

\bibitem{dip}
V.~Berezinsky, A.~Z.~Gazizov and S.~I.~Grigorieva,
Phys.\ Rev.\  D {\bf 74}, 043005 (2006) [hep-ph/0204357];


\bibitem{Roth:2007in}
  M.~Roth,
  arXiv:0706.2096 [astro-ph].




\bibitem{Razzaque:2002kb}
  S.~Razzaque {\it et al.},
  Phys.\ Rev.\ Lett.\  {\bf 90} (2003) 241103.
  S.~R.~Kelner {\it et al.},
  Phys.\ Rev.\  D {\bf 74} (2006) 034018.




\bibitem{Mucke:1999yb}
  A.~Mucke {\it et al.},
  Comput.\ Phys.\ Commun.\  {\bf 124} (2000) 290;
  A.~Mucke {\it et al.},
  astro-ph/9905153.
  A.~Mucke,{\it et al.},
  Publ.\ Astron.\ Soc.\ Austral.\  {\bf 16} (1999) 160


\bibitem{Lipari:2007su}
  P.~Lipari, M.~Lusignoli and D.~Meloni,
  Phys.\ Rev.\  D {\bf 75} (2007) 123005.

\bibitem{Chiaberge:2001ek}
  M.~Chiaberge, A.~Capetti and A.~Celotti,
  Mon.\ Not.\ Roy.\ Astron.\ Soc.\  {\bf 324} (2001) L33;
  J.~P.~Lenain {\it et al.},
  arXiv:0710.2847 [astro-ph].

\bibitem{Aharonian:2005ar}
  F.~Aharonian {\it et al.}  [H.E.S.S. Collaboration],
  Astron.\ Astrophys.\  {\bf 441} (2005) 465.

\bibitem{Kabuki:2007am}
  S.~Kabuki {\it et al.}  [CANGAROO-III Collaboration],
  arXiv:0706.0367 [astro-ph].


\bibitem{Hardcastle:2003ye}
  M.~J.~Hardcastle {\it et al.},
  Astrophys.\ J.\  {\bf 593} (2003) 169.

\bibitem{Horiuchi:2005kz}
  S.~Horiuchi {\it et al.},
  Publ.\ Astron.\ Soc.\ Jap.\  {\bf 58} (2006) 211.



\bibitem{Rachen:1998fd}
  J.~P.~Rachen and P.~Meszaros,
  Phys.\ Rev.\  D {\bf 58} (1998) 123005.



\bibitem{Steinle:1998yr}
  H.~Steinle {\it et al.},
  Astron.\ Astrophys.\ {\bf 330} (1998) 97-107.

\bibitem{Kinzer:1995yr}
  R.~L.~Kinzer {\it et al.},
  Astrophys.\ J.\ {\bf 449} (1995) 105-118.


\bibitem{Sreekumar:1999xw}
  P.~Sreekumar, {\it et al.}, 
  Astropart.\ Phys.\  {\bf 11} (1999) 221


\bibitem{Kachelriess:2008qx}
  M.~Kachelriess, S.~Ostapchenko and R.~Tomas,
  arXiv:0805.2608 [astro-ph].



\bibitem{Gandhi:1998ri}
  R.~Gandhi {\it et al.},
  Phys.\ Rev.\  D {\bf 58} (1998) 093009.


\bibitem{Anchordoqui:2004eb}
  L.~A.~Anchordoqui {\it et al.},
  Phys.\ Lett.\  B {\bf 621} (2005) 18.


\bibitem{Serpico:2005sz}
  P.~D.~Serpico and M.~Kachelriess,
  Phys.\ Rev.\ Lett.\  {\bf 94} (2005) 211102;
  W.~Winter,
  Phys.\ Rev.\  D {\bf 74} (2006) 033015;
  P.~D.~Serpico,
  Phys.\ Rev.\  D {\bf 73} (2006) 047301.

\bibitem{Gorbunov:2007ja}
  D.~Gorbunov {\it et al.}, 
  arXiv:0711.4060 [astro-ph].



\end{thebibliography}
\end{document}